\documentclass{article}

\PassOptionsToPackage{numbers}{natbib}

\usepackage[final]{neurips_2023}

\usepackage[utf8]{inputenc} 
\usepackage[T1]{fontenc}    
\usepackage{hyperref}       
\usepackage{url}            
\usepackage{booktabs}       
\usepackage{amsfonts}       
\usepackage{nicefrac}       
\usepackage{microtype}      
\usepackage{xcolor}         

\usepackage[inkscapelatex=false]{svg} 

\usepackage{setspace}

\title{A collection of principles for guiding and \\evaluating large language models}

\author{
  Konstantin Hebenstreit \\
  Institute of Artificial Intelligence \\
  Medical University of Vienna \\
  1090 Vienna, Austria \\
  \And
  Robert Praas \\
  KTH Royal Institute of Technology \\
  114 28 Stockholm, Sweden \\
  \And
  Matthias Samwald \\
  Institute of Artificial Intelligence \\
  Medical University of Vienna \\
  1090 Vienna, Austria \\
  \texttt{matthias.samwald [@] meduniwien.ac.at} \\
}

\begin{document}

\maketitle

\begin{abstract}
Large language models (LLMs) demonstrate outstanding capabilities, but challenges remain regarding their ability to solve complex reasoning tasks, as well as their transparency, robustness, truthfulness, and ethical alignment. In this preliminary study, we compile a set of core principles for steering and evaluating the reasoning of LLMs by curating literature from several relevant strands of work: structured reasoning in LLMs, self-evaluation/self-reflection, explainability, AI system safety/security, guidelines for human critical thinking, and ethical/regulatory guidelines for AI. We identify and curate a list of 220 principles from literature, and derive a set of 37 core principles organized into seven categories: assumptions and perspectives, reasoning, information and evidence, robustness and security, ethics, utility, and implications. We conduct a small-scale expert survey, eliciting the subjective importance experts assign to different principles and lay out avenues for future work beyond our preliminary results. We envision that the development of a shared model of principles can serve multiple purposes: monitoring and steering models at inference time, improving model behavior during training, and guiding human evaluation of model reasoning.
\end{abstract}

\section{Background}
\setstretch{1.125}

Large language models (LLMs) such as GPT-4 \citep{openai_2023}, Flan-PaLM \citep{chung_2022} and Claude \citep{bai_2022a} demonstrate outstanding abilities on a wide variety of tasks. However, LLMs still encounter significant challenges when addressing complex reasoning tasks \citep{ye_other_2022, lu_2023, livin_2022, ruis_2022, jones_2022, savelka_2023}, and there are substantial concerns regarding their truthfulness, transparency, robustness and alignment with ethical values \citep{creswell_2022, wang_other_2022, wolf_2023, liang_2022, evans_2021}. These limitations hinder the applicability of LLMs in complex and critical domains, such as automating scientific research or augmenting medical decision-making.

\begin{figure}[h]
    \centerline{\includegraphics[width=0.95\columnwidth]{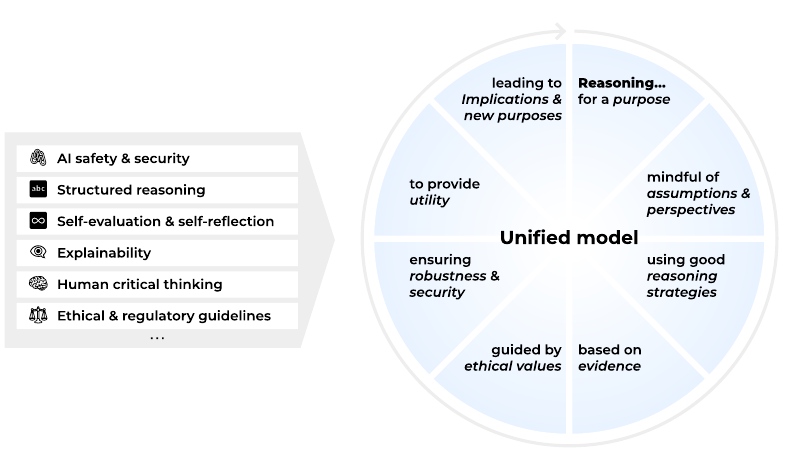}}
    \caption{We aim to integrate insights from multiple distinct strands of work into a set of core principles guiding reasoning in LLM-based AI systems.}
    \label{fig-circle}
\end{figure}

In this work, we review literature from six distinct fields of research to establish a conceptual model of high-level principles for guiding the reasoning and responses of LLMs (Figure \ref{fig-circle}). 

Recent work on \textbf{structured reasoning} was initiated by the finding that LLMs can generate ‘chains of thought’ that improve reasoning performance on complex tasks and improve the transparency of reasoning processes \cite{wei_2022, kojima_2022}. Succeeding work aimed at further improving reasoning performance and interpretability by effectively decomposing the reasoning process and, optionally, interleaving reasoning with LLM-guided access to external tools such as information retrieval \cite{zhou_2022, press_2022, yao_2022, jung_2022, dua_2022, trivedi_2022, khattab_2022, paranjape_2023, lu_2023a, shinn_2023, kim_2023, nair_2023}.

The increased capacity of generating reasoning chains and long-form textual answers led to varied considerations and approaches towards their evaluation by human annotators \cite{livin_2022, ott_2023, kung_2023, singhal_2023}. Given the limited scalability of human evaluation, research on utilizing LLM models for automated \textbf{self-evaluation and self-reflection} in LLM models emerged. This work was initially focused on issues such as bias and toxicity by identifying offensive content \cite{perez_2022} and trying to enable ‘self-debiasing’ \cite{schick_2021}. More recent work on reinforcement learning from AI feedback \cite{bai_2022a} devised evaluation instructions that form a ‘constitution’ of norms for LLMs. By creating a preference model for reinforcement learning, a highly capable and value-aligned LLM was trained with minimal human feedback. This approach was recently replicated \cite{sun_2023}. Further work focused on the evaluation of reasoning chains by LLMs \cite{golovneva_2022}, and structured interaction between multiple instances of LLMs \cite{nair_2023, li_2023}.

Work on AI system \textbf{explainability} strives to understand how AI systems can be made more transparent and how the explanatory utility of their output can be increased. Concerns associated with explainability include practical utility, faithfulness, contrastiveness, simulatability, and causality \cite{bai_2022, bai_2022a, ganguli_other_2022, jacovi_2020, vilone_2020, wiegreffe_2021}.

AI system \textbf{security and safety} is concerned with making systems robust against misinformation or adversarial attacks \cite{kojima_2022, kuhn_2022}, as well as creating AI systems that do not deceive or cause harm \cite{bai_2022, bai_2022a, hendrycks_2021}.

Outside of the realm of computer science, other disciplines like philosophy and psychology are of relevance, providing general principles guiding \textbf{human critical thinking} \cite{elder_2019} that can be applied to LLMs.

Finally, a growing body of research and regulations around \textbf{ethical and regulatory guidelines} for AI system creation and deployment is emerging, exemplified by the current draft of the EU AI act or the OECD AI principles \cite{jobin_2019, buruk_2020, europeancommission_website_2021, oecd_website_2020}.

We build on these preceding strands of work to derive a unified set of core principles for guiding and evaluating the reasoning of LLMs.

\section{Methods}

We conducted an extensive literature review in the domains outlined above, extracting relevant concerns, system instructions, and guidelines. We compiled the results into a table of guiding principles and categorized them into a small number of overarching categories (e.g. utility, ethics, robustness and security).

Through iterative revisions, we distilled a small, consolidated set of core principles that incorporated the majority of concerns relevant to current LLMs. We aimed to ensure that principles in this set were formulated as broadly as possible, thus facilitating their application across a diverse range of reasoning tasks and enhancing their adaptability to multimodal setups. 

For external validation, we identified domain experts by compiling a list of authors from recent papers on LLM reasoning, LLM evaluation, or AI ethics. We strived to contact experts from a variety of academic and industry institutions,  and asked them to fill out an anonymous survey judging the set of core principles. The experts were asked to imagine evaluating an LLM response in a critical application domain and to rate the importance of each principle on a 5-point Likert scale spanning from ‘not important’ to ‘very important’. A screenshot of the web survey is shown in supplementary material S1.

\section{Results}

\begin{figure}[h!]
    \centerline{\includegraphics[width=1\columnwidth]{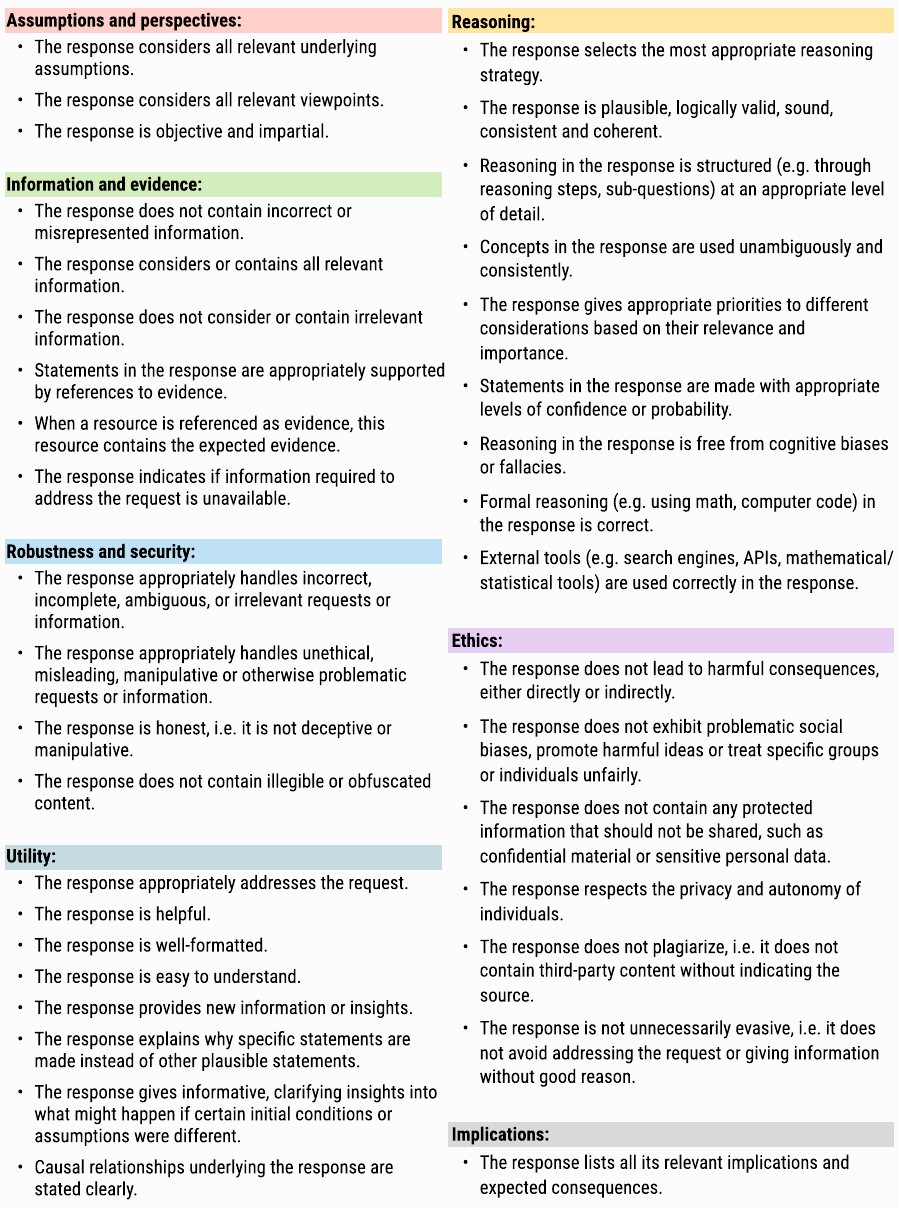}}
    \caption{The set of core principles compiled from literature.}
    \label{fig-core-principles}
\end{figure}

The literature review yielded a total of thirty documents from which we extracted 220 principles that we categorized into eight categories. From this collection, we derived a core set containing 37 principles. The distribution of principles across categories is detailed in Table 1, and the resultant core principles are listed in Figure \ref{fig-core-principles}. The table of raw results from the literature review is available online.\footnote{\href{https://docs.google.com/spreadsheets/d/1tuml5PUL-uAExBukCZ741JXu3Sze6djnsErm5k3yqzU/}{https://docs.google.com/spreadsheets/d/1tuml5PUL-uAExBukCZ741JXu3Sze6djnsErm5k3yqzU/}} 


\begin{table}[h!]
  \caption{Number of principles across categories}
  \centering
  \begin{tabular}{lcc}
    \toprule
    \textbf{Category}            & \textbf{Collected from literature}& \textbf{Derived core set} \\
    \midrule
    Assumptions and perspectives & 28                         & 3                        \\
    Reasoning                    & 40                         & 9                        \\
    Information and evidence     & 18                         & 6                        \\
    Ethics                       & 79                         & 6                        \\
    Robustness and security      & 11                         & 4                        \\
    Utility                      & 30                         & 8                        \\
    Implications                 & 2                          & 1                        \\
    General                      & 12                         & —                        \\
    \midrule
    Total                        & 220                        & 37                       \\                         
    \bottomrule
  \end{tabular}
\end{table}

We contacted 66 domain experts for external validation, of which 7 filled out the anonymous survey. Results of the survey are detailed in S2 Appendix. For the majority of principles, median responses were 'important' or 'very important'. However, the span of responses for some principles was wide, sometimes spanning from 'not important' to 'very important' for a single item.

\newpage
\clearpage
\section{Discussion}

We envision that sets of principles like the one presented here can serve multiple purposes, such as steering models at inference time, e.g., through formulating principles as instructions that structure reasoning processes; monitoring, critiquing, and improving model behavior, e.g., through automated self-critique and improvement at inference time (self-reflection) or during training (‘constitutional AI’); and assisting human evaluation of the reasoning process in model responses. The set of core principles listed here can be adapted, shortened, expanded, and re-formulated to fit such a variety of application scenarios.

We expect the social impacts of our work to be primarily beneficial, as the principles are designed to support ethical, robust, and secure reasoning in LLMs. This also reflects the sources of the different principles (explainability, safety and security, and ethical regulations). However, there is significant uncertainty about how well instructing and training advanced models with such principles can ensure that such models are actually being adhered to. The approach outlined here therefore must be seen as only one aspect of a comprehensive strategy for ensuring model alignment.

Our literature review did not cover all relevant existing work on the principles of reasoning. However, we observed a significant level of redundancy in the concerns that could be gathered from additional literature and expect that our current literature review has addressed the majority of concerns at our targeted level of generality. 

While we strove to formulate the core principles as universally as possible, they are nonetheless biased towards language-based interactions enabled by the current generation of LLMs, such as question-answering systems or general-purpose dialog models. We anticipate adding concerns and objectives that will emerge as AI systems increasingly function as autonomous agents with a wide variety of capabilities and potential risks.

The current work is preliminary, and the set of core principles is not exhaustive. While we conducted an extensive literature review as the basis of our work, we derived the eventual set based on subjective judgments of relevance to current and emerging AI systems. The small-scale expert survey serves as a starting point for more elaborate community processes for eliciting relevant principles. This could happen as a closed expert-based process, such as a multi-round Delphi study, or as an open online discussion platform, on which principles can be created, rated, and updated iteratively. Future work should also increase emphasis on gathering input from a broader set of perspectives, including industry stakeholders. We invite the community to provide feedback and further ideas. 

Finally, the practical utility of specific core principles and the set of principles as a whole should be explored through empirical research, e.g., by deriving prompts for state-of-the-art language models and evaluating their influence on system performance, explainability, and ethical alignment.

\begin{ack}
We thank Nathalie Kirch for providing guidance in designing the questionnaire and Louis Kiesewetter for providing feedback on the manuscript.


\end{ack}

\section{Supplementary Material}

\paragraph*{S1.}
Survey screenshot

\paragraph*{S2.}
Survey results


\medskip

\newpage
\setstretch{1.0}
\bibliography{P4.bib}


\newpage
\clearpage

\section*{S1: Survey screenshot}

\begin{figure}[h!]
    \centering
    \includegraphics[width=1\linewidth]{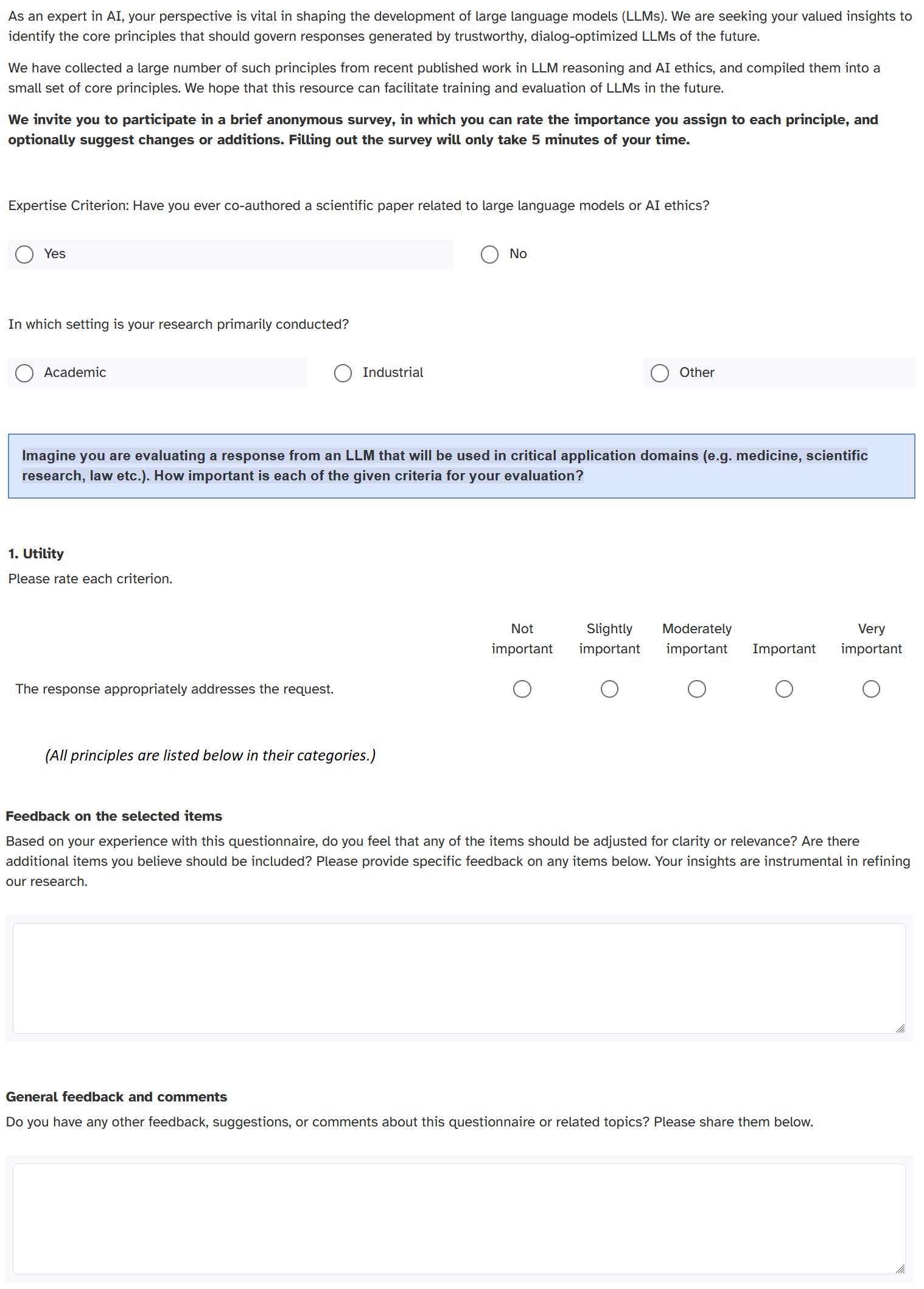}
    \caption{Screenshot of the survey page. Only the first item representing a principle is displayed.}
    \label{fig-survey-screenshot}
\end{figure}

\newpage
\clearpage

\section*{S2: Survey results}
\begin{table}[h]
\tiny
\centering
\caption{Expert survey results with minimum, median and maximum importance ratings (N=7).\\ 1: not important, 2: slightly important, 3: moderately important, 4: important, 5: very important. }
\begin{tabular}{p{0.9\linewidth}ccc}
\toprule
\textbf{Text} & \textbf{Min} & \textbf{Median} & \textbf{Max} \\
\midrule
\multicolumn{4}{l}{\textbf{Assumptions and perspectives:}} \\
The response considers all relevant underlying assumptions. & 3 & 4 & 5 \\
The response considers all relevant viewpoints. & 2 & 4 & 5 \\
The response is objective and impartial. & 2 & 4 & 5 \\
\addlinespace
\multicolumn{4}{l}{\textbf{Reasoning:}} \\
The response selects the most appropriate reasoning strategy. & 2 & 4 & 4 \\
The response is plausible, logically valid, sound, consistent and coherent. & 4 & 5 & 5 \\
Reasoning in the response is structured (e.g. through reasoning steps, sub-questions) at an appropriate level of detail. & 4 & 4 & 5 \\
Concepts in the response are used unambiguously and consistently. & 3 & 4 & 5 \\
The response gives appropriate priorities to different considerations based on their relevance and importance. & 3 & 4 & 5 \\
Statements in the response are made with appropriate levels of confidence or probability. & 2 & 4 & 5 \\
Reasoning in the response is free from cognitive biases or fallacies. & 2 & 5 & 5 \\
Formal reasoning (e.g. using math, computer code) in the response is correct. & 3 & 5 & 5 \\
External tools (e.g. search engines, APIs, mathematical/statistical tools) are used correctly in the response. & 3 & 5 & 5 \\
\addlinespace
\multicolumn{4}{l}{\textbf{Information and evidence:}} \\
The response does not contain incorrect or misrepresented information. & 4 & 5 & 5 \\
The response considers or contains all relevant information. & 3 & 4 & 5 \\
The response does not consider or contain irrelevant information. & 2 & 4 & 5 \\
Statements in the response are appropriately supported by references to evidence. & 3 & 4 & 5 \\
When a resource is referenced as evidence, this resource contains the expected evidence. & 3 & 5 & 5 \\
The response indicates if information required to address the request is unavailable. & 3 & 4 & 5 \\
\addlinespace
\multicolumn{4}{l}{\textbf{Robustness and security:}} \\
The response appropriately handles incorrect, incomplete, ambiguous, or irrelevant requests or information. & 4 & 5 & 5 \\
The response appropriately handles unethical, misleading, manipulative or otherwise problematic requests or information. & 4 & 5 & 5 \\
The response is honest, i.e. it is not deceptive or manipulative. & 4 & 5 & 5 \\
The response does not contain illegible or obfuscated content. & 3 & 4 & 5 \\
\addlinespace
\multicolumn{4}{l}{\textbf{Ethics:}} \\
The response does not lead to harmful consequences, either directly or indirectly. & 3 & 5 & 5 \\
The response does not exhibit problematic social biases, promote harmful ideas or treat specific groups or individuals unfairly. & 4 & 5 & 5 \\
The response does not contain any protected information that should not be shared, such as confidential material or sensitive personal data. & 4 & 5 & 5 \\
The response respects the privacy and autonomy of individuals. & 4 & 5 & 5 \\
The response does not plagiarize, i.e. it does not contain third-party content without indicating the source. & 2 & 3 & 5 \\
The response is not unnecessarily evasive, i.e. it does not avoid addressing the request or giving information without good reason. & 1 & 4 & 5 \\
\addlinespace
\multicolumn{4}{l}{\textbf{Utility:}} \\
The response appropriately addresses the request. & 5 & 5 & 5 \\
The response is helpful. & 4 & 5 & 5 \\
The response is well-formatted. & 3 & 4 & 5 \\
The response is easy to understand. & 4 & 4 & 5 \\
The response provides new information or insights. & 1 & 4 & 5 \\
The response explains why specific statements are made instead of other plausible statements. & 3 & 4 & 5 \\
The response gives informative, clarifying insights into what might happen if certain initial conditions or assumptions were different. & 2 & 4 & 5 \\
Causal relationships underlying the response are stated clearly. & 1 & 4 & 5 \\
\addlinespace
\multicolumn{4}{l}{\textbf{Implications:}} \\
The response lists all its relevant implications and expected consequences. & 3 & 4 & 4 \\
\bottomrule
\end{tabular}
\end{table}
Out of seven respondents, six listed their background as 'academia' and one listed their background as 'industry'.

\end{document}